\documentclass[10pt,journal,compsoc]{IEEEtran}

\usepackage[T1]{fontenc}
\usepackage{cite}
\usepackage{amsmath,amssymb}
\usepackage{graphicx}
\usepackage{booktabs}
\usepackage{makecell}
\usepackage{multirow}
\usepackage{tabularx}
\usepackage{array}
\usepackage{url}
\usepackage{xcolor}
\usepackage{microtype}
\usepackage[hidelinks]{hyperref}

\graphicspath{{figures/}}

\newcolumntype{Y}{>{\raggedright\arraybackslash}X}
\newcolumntype{C}{>{\centering\arraybackslash}X}
\newcolumntype{M}[1]{>{\centering\arraybackslash}m{#1}}
\newcolumntype{L}[1]{>{\raggedright\arraybackslash}m{#1}}

\begin{document}

\title{{PLC-Bin2Src: Retrieving Corresponding Structured Text Source Files for PLC Binaries}}

\author{{Ang Jia, Yaxin Duan, He Jiang, Ming Fan, Zhilei Ren, and Xiaochen Li}%
  \thanks{{Corresponding author: He Jiang. E-mail: \texttt{jianghe@dlut.edu.cn}.}}%
  \thanks{{Ang Jia, Yaxin Duan, He Jiang, Zhilei Ren, and Xiaochen Li are with the School of Software, Dalian University of Technology, Dalian, China.}}%
  \thanks{{Ming Fan is with the School of Cyber Science and Engineering, Xi'an Jiaotong University, Xi'an, China.}}%
}

\IEEEtitleabstractindextext{%
\begin{abstract}
PLC developers reuse existing components and third-party libraries to reduce the cost of developing new applications, but this reuse can also introduce security risks. When components are distributed only in binary form, provenance tracing, vulnerability assessment, and security auditing become more difficult. To support these analyses, binary2source matching can retrieve the corresponding Structured Text (ST) source file for a PLC binary from a candidate source repository. However, due to the properties of PLC software, this task {faces} {three} {challenges:} cross-platform compilation heterogeneity, {a} binary--source representation gap, and {inadequate representations of PLC semantics}. Therefore, {we} {present} PLC-Bin2Src, a cross-platform binary2source matching framework {that} {retrieves} corresponding ST source files for binaries produced by CODESYS, GEB, OpenPLC v2, and OpenPLC v3. First, PLC-Bin2Src {uses} {platform-aware} binary frontends to recover user control logic across toolchains{.} Then, ST-to-C conversion and shared normalization are used to reduce {syntactic differences between source- and binary-side representations}{.} Next, {control--data flow graphs (CDFGs) and function call graphs (FCGs) are constructed to capture program semantics, while recovered symbols provide complementary identity evidence.} Finally, PLC-Bin2Src {compares these representations and combines their similarity scores} to rank source candidates. We evaluate PLC-Bin2Src {on} PLC-BEAD{.} The results show that PLC-Bin2Src achieves 95.89\% Recall@1, 99.66\% Recall@5, and an MRR of 0.9769 across four PLC platforms.
\end{abstract}

}

\maketitle

\section{Introduction}

\IEEEPARstart{P}{rogrammable} logic controllers (PLCs) are core execution components of industrial automation systems. {Deployed PLC programs control physical processes by} cyclically reading field inputs, executing control logic, and updating actuators~\cite{iec61131,keliris2019icsref}. PLC developers reuse existing components and third-party libraries to reduce development costs, but this reuse can also introduce security risks. Moreover, some reused libraries are delivered only in compiled form. For example, CODESYS supports {\texttt{compiled-library}} packages for library distribution.
{Downstream} developers can integrate {these} packages without access to their source code or implementation details~\cite{codesysCompiledLibraries,codesysLibraryDeployment}. {This} binary-only reuse makes it difficult for analysts to {identify} {the incorporated} components, {their} logic, and {their} {origins}.

Software composition analysis addresses this visibility problem by identifying reused components and linking them to known source implementations. {Existing} PLC binary reverse engineering frameworks, such as ICSREF~\cite{keliris2019icsref} and PLC-BinX~\cite{jia2026plcbinx}, focus on reconstructing executable structure and control semantics from proprietary artifacts. They do not identify which source file corresponds to a binary artifact. {Binary2source} matching {is} {therefore} needed to retrieve the corresponding source implementations {and support} {the creation of dependency inventories}, vulnerability assessment, provenance tracing, and security auditing~\cite{duan2017osspolice,miyani2017binpro,jiang2024binaryai}.

PLC software is commonly written in Structured Text (ST) {and compiled into} executable artifacts by platform-specific toolchains.
Fig.~\ref{fig:cross_platform_acosh} illustrates the cross-platform representation gap and the {binary--source representation gap} using \texttt{ACOSH.ST} from PLC-BEAD~\cite{achamyeleh2025plcbead}. Binaries generated from this ST program yield three different ARM-level views: OpenPLC v3 implements {\texttt{SQRT}} and {\texttt{LN}} through calls to type-specific runtime functions; CODESYS v3 compiles \texttt{SQRT} to a floating-point square-root instruction and invokes \texttt{REAL32\_\_LN} indirectly; and GEB {calls} \texttt{gebx} math {functions} and \texttt{geb\_hook}. {Even} with the {same} instruction set architecture, instruction context, call targets, and recovered {function} boundaries differ across platforms{.} {These binary views also} share little surface syntax with ST. Therefore, {binary2source} matching for PLC software {faces} three challenges:

\begin{figure*}[t]
	\centering
	\includegraphics[width=0.96\textwidth]{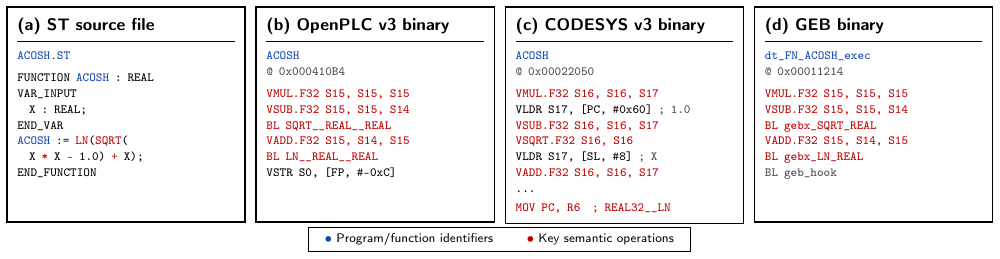}
	\caption{Four views of \texttt{ACOSH.ST}: (a) ST source; (b) OpenPLC v3 ARM disassembly; (c) CODESYS v3 ARM disassembly; and (d) GEB ARM disassembly.}
	\label{fig:cross_platform_acosh}
\end{figure*}

\begin{enumerate}
	\item {\emph{Cross-platform compilation heterogeneity.} PLC platforms differ in binary formats, compilation toolchains, and runtime organization, so the same ST logic can yield different function boundaries, instruction patterns, and call structures.}
	\item {\emph{Binary--source representation gap.} ST expresses control logic through typed variables, structured statements, and function calls, whereas binaries encode it as machine instructions, memory accesses, and runtime calls. Compilation can obscure variable identities and type information, reorganize expressions, and introduce temporary values and explicit conversions. {Corresponding} computations and data accesses can {therefore} have substantially different surface forms, making direct comparison unreliable.}
	\item {\emph{Inadequate representations {of} PLC semantics.} PLC programs can contain similar operations yet implement different control logic, since the role of each operation depends on its surrounding context. Representations based on isolated operations or local patterns may therefore fail to distinguish semantically different source candidates. {Incomplete binary-level evidence makes} these distinctions {harder to capture} reliably.}
\end{enumerate}



{We} propose {PLC-Bin2Src,} a binary2source matching framework for PLC binaries{, to address these challenges}. First, {platform-aware binary frontends} address cross-platform compilation heterogeneity {by} {recovering} user control logic through toolchain-specific disassembly, core-function extraction, and decompilation. Second, {an ST-to-C source frontend expresses source control logic in C} to {help} bridge the binary--source representation gap{.} {Shared} normalization aligns variable and memory-access forms, type conversions, and platform-specific operation and library-call names {across the source- and binary-side representations}. Third, {PLC-Bin2Src} {constructs} {three complementary program representations: control--data flow graphs (CDFGs)} for {intraprocedural} control--data dependencies, {function call graphs (FCGs)} for {interprocedural} call relationships, and recovered symbols for complementary identity evidence. Finally, {PLC-Bin2Src converts the features of each representation into file-level vectors, compares the corresponding binary and source vectors, and combines their similarity scores to rank source candidates.}

We evaluate PLC-Bin2Src on PLC-BEAD~\cite{achamyeleh2025plcbead}. {PLC-Bin2Src} {achieves} 95.89\% Recall@1, 99.66\% Recall@5, and an MRR of 0.9769 when retrieving corresponding ST source files for PLC binaries compiled {for} four platforms: CODESYS, GEB, OpenPLC v2, and OpenPLC v3.

{This} paper {makes} {the} {following contributions}:

\begin{itemize}
	\item To the best of our knowledge, PLC-Bin2Src is the first binary2source matching framework for retrieving {corresponding ST source files} for PLC binaries.
	\item {PLC-Bin2Src provides a unified framework for aligning PLC binaries with ST source code. Platform-aware binary frontends, ST-to-C conversion, and shared normalization bring both sides into comparable C-like forms for subsequent representation construction and matching.}
	\item PLC-Bin2Src constructs three complementary program representations {(}CDFGs, FCGs, and recovered symbols{)} and combines their similarity scores to rank candidate ST files.
	\item We evaluate PLC-Bin2Src on PLC-BEAD. {PLC-Bin2Src} achieves 95.89\% Recall@1, 99.66\% Recall@5, and an MRR of 0.9769 across the four platforms.
\end{itemize}

\section{Background}

\subsection{PLC Program Model and Compilation Process}

PLCs typically execute control tasks {through} a cyclic scan mechanism {that} {reads} field inputs, {runs} user control logic, and {updates} outputs~\cite{lopez2024sok}. The IEC~61131-3 standard specifies PLC programming languages and their program organization. ST is a textual language suitable for expressing conditional decisions, loops, numerical operations, and function calls. An ST program consists of program organization units (POUs), including \texttt{PROGRAM}, \texttt{FUNCTION}, and \texttt{FUNCTION\_BLOCK}, and {these} POUs jointly describe control logic through variable access and function calls~\cite{iec61131}.

PLC source code must be transformed by a platform compilation toolchain into a binary that can execute on the controller. In OpenPLC, for example, MATIEC translates an IEC~61131-3 program into equivalent C code, which a native compiler then converts into an executable for the target platform~\cite{matiec}. Compilation introduces wrapper functions, cyclic scheduling, variable-access interfaces, temporary values, and runtime support logic. Disassembly can recover low-level arithmetic and calls, but source logic remains distributed across user functions, wrappers, and the cyclic execution chain; this semantic mismatch motivates binary2source matching.

\subsection{Cross-Platform Differences in PLC Binaries}

PLC platforms vary in binary format, target architecture, compilation toolchain, runtime organization, and symbol-naming rules~\cite{jia2026plcbinx,benkraouda2023plcspecific,achamyeleh2025plcbead}. Fig.~\ref{fig:cross_platform_acosh} illustrates these differences using the same \texttt{ACOSH.ST} program. Panel (a) expresses the computation as a nested ST expression. In panel (b), OpenPLC v3 exposes ARM floating-point instructions together with typed mathematical runtime calls. Panel (c) shows the CODESYS v3 \texttt{ACOSH} code area: multiplication, subtraction, square root, and addition remain visible as floating-point instructions, while \texttt{REAL32\_\_LN} is reached through an indirect call. Panel (d) shows that GEB retains related floating-point operations but uses \texttt{gebx} runtime calls and an additional hook. Thus, the same source program yields different function contexts, call targets, memory-access forms, and recoverable semantic units across platforms.

The comparison also shows that surface-level similarity is not a reliable cross-platform signal. OpenPLC uses typed runtime calls, GEB uses \texttt{gebx} calls and a hook, and CODESYS combines a hardware square-root instruction with an indirect logarithm call. Binary2source matching must therefore localize the code regions that carry user control logic and normalize operations, calls, and identifiers from platform-specific representations before comparing them with ST source files.

\section{Method}

\begin{figure*}[!t]
	\centering
	\includegraphics[width=\textwidth]{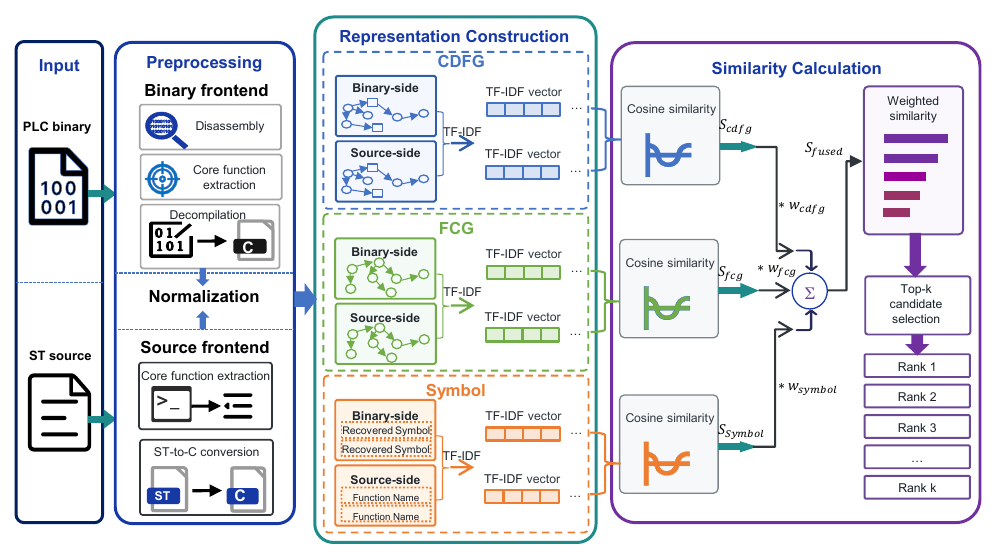}
	\caption{Overview of the PLC-Bin2Src workflow.}
	\label{fig:framework}
\end{figure*}

PLC-Bin2Src models binary2source matching as a source-candidate ranking problem. Fig.~\ref{fig:framework} summarizes its three stages: preprocessing, representation construction, and similarity calculation. The preprocessing stage recovers binary-side core functions, transforms source-side ST programs, and normalizes both representations. The representation-construction stage builds control--data flow graph (CDFG), function call graph (FCG), and recovered-symbol representations and converts their features into term frequency--inverse document frequency (TF-IDF) vectors. The figure shows the three representations in blue, green, and orange, respectively. The similarity-calculation stage computes cosine similarities separately for the three representations, combines them with fixed equal weights, ranks every compatible ST file, and returns the top-$k$ candidates. In the figure, $S_{\mathrm{cdfg}}$, $S_{\mathrm{fcg}}$, and $S_{\mathrm{symbol}}$ denote the similarity scores for the three representations; the corresponding $w$ terms are all fixed at $1/3$; and $S_{\mathrm{fused}}$ denotes the resulting score used for ranking.


\subsection{Preprocessing}

\begin{figure}[!t]
	\centering
	\includegraphics[width=\columnwidth]{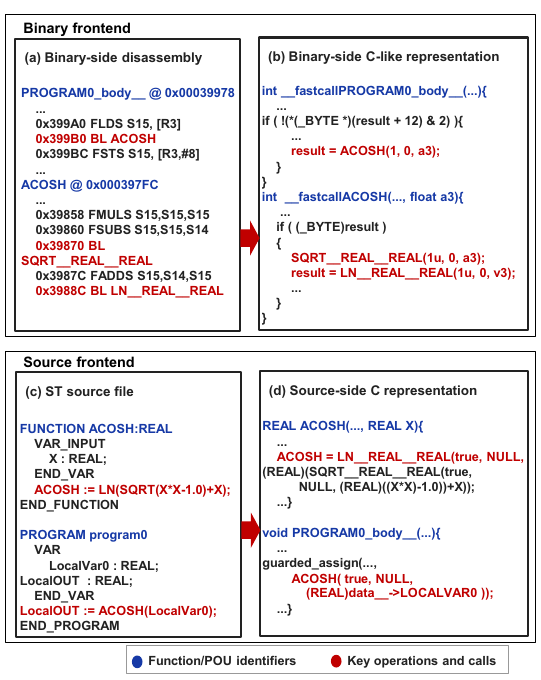}
	\caption{{Binary-side and source-side} frontend processing for \texttt{ACOSH}.}
	\label{fig:frontend_example}
\end{figure}

Fig.~\ref{fig:frontend_example} illustrates how the two frontends transform \texttt{ACOSH} into comparable representations. {The binary frontend extracts and decompiles core functions to obtain a C-like representation, while the source frontend converts the ST program into a C representation.} In the figure, the upper and lower parts show {binary-side and source-side} processing, respectively; blue identifies program units and core functions, red highlights operations and calls traceable across representations, and ellipses indicate omitted auxiliary code.

\subsubsection{Binary Frontend}

The binary frontend extracts function-level representations related to user control logic from binaries produced by different PLC platforms. It comprises disassembly, core-function extraction, and decompilation. The {upper} part of Fig.~\ref{fig:frontend_example} shows the resulting representation for \texttt{ACOSH}.

\textbf{Disassembly:} The system parses deployment artifacts and recovers the function identifiers, address ranges, and program structures required for core-function extraction. For OpenPLC and GEB ELF files, it obtains function boundaries and call relationships from standard binary structures and disassembly. A CODESYS \texttt{.app} file is a proprietary application container without the header and section table of a standard ELF file, so it requires a different process. For CODESYS, the binary frontend {obtains} recovered executable ranges, call relationships, and metadata-linked POU names as inputs. The resulting boundaries, address ranges, and POU-name clues support subsequent CODESYS core-function extraction.

\textbf{Core-function extraction:} {Using} the recovered information, PLC-Bin2Src applies platform-specific rules to select core functions associated with user control logic. For OpenPLC, it retains the cyclic-execution wrapper and user functions. For GEB, it selects program-, function-, and function-block execution routines according to the \texttt{dt\_PR\_*\_exec}, \texttt{dt\_FN\_*\_exec}, and \texttt{dt\_FB\_*\_exec} naming patterns. For CODESYS, it associates metadata-linked entry pointers with recovered function ranges and uses function calls to identify POU execution candidates while reducing the priority of common templates and runtime routines. The extracted core functions are then passed to the decompilation stage.

\textbf{Decompilation:} After core-function extraction, PLC-Bin2Src generates C-like representations for all platforms. For OpenPLC and GEB, target functions are selected from ELF function information and directly decompiled. For CODESYS, POU execution candidates are first recovered from function boundaries, recovered symbols, and call information. Then, the address ranges of {selected} POU functions {are decompiled} into C-like code. The call relations between POU roots and helper functions are still {preserved}.

\subsubsection{Source Frontend}

The source frontend transforms ST programs into function-level representations {comparable} {to} the binary-side representations. PLC-Bin2Src does not directly match raw ST text, but instead performs two steps in sequence: ST-to-C conversion and core-function extraction. The {lower} part of Fig.~\ref{fig:frontend_example} shows the conversion of \texttt{ACOSH} from ST source code to a C representation.

\textbf{ST-to-C conversion:} For each ST source sample, PLC-Bin2Src reads the source program and uses the MATIEC compiler~\cite{matiec} to generate a C-style representation. It then uses a C preprocessor to resolve macros, variable-access wrappers, and conditional-compilation structures in the generated code, obtaining effective C code after macro replacement and conditional selection.

\textbf{Core-function extraction:} From the generated C-like representation, the system extracts functions that represent user control logic. It excludes \texttt{program0}, initialization routines, and platform-support functions such as \texttt{copy}, \texttt{print}, and \texttt{dump}; it prioritizes the function or function-block body whose name corresponds to the current sample. When name clues are insufficient, the \texttt{*\_body\_\_} body generated by the compilation toolchain is retained as a candidate, keeping the source-side representation focused on user control logic.

\subsubsection{Normalization}

The source-side C representation and the binary-side C-like representation differ in their naming conventions, memory-access forms, and expressions of equivalent operations. PLC-Bin2Src reduces these differences through normalization rules for operation names, variable and memory-access forms, and numerical literals. These rules standardize both common code forms and the labels used in subsequent representation construction.

\textbf{Operation normalization:}
Different PLC toolchains can implement the same mathematical operation through differently named runtime functions. PLC-Bin2Src removes recognized compiler-generated decorations and type suffixes from call names and maps known platform-specific interfaces to common operation labels. For example, \texttt{SQRT\_\_REAL\_\_REAL} and \texttt{gebx\_SQRT\_REAL} are both associated with \texttt{SQRT}. Recognized primitive-operation calls are also mapped to the corresponding arithmetic or logical operator categories. This allows operations expressed through different calling conventions and naming schemes to contribute comparable features. On the GEB binary side, runtime-hook calls such as \texttt{geb\_hook} are filtered to reduce features introduced by runtime support code.

\textbf{Variable and memory-access normalization:}
Source-side code often accesses PLC variables through named fields, whereas binary-side C-like code may express accesses through temporary variables and pointer offsets. PLC-Bin2Src abstracts local-variable names in structural features while retaining their occurrences for local definition--use analysis. It also rewrites recognized memory-access patterns into common slot or memory labels. Explicit source-side \texttt{.value} and \texttt{.flags} fields retain their value and flag roles. On the binary side, access widths and recurring offset patterns provide heuristic cues for assigning these roles; other recognized accesses receive generic memory labels. These abstractions reduce dependence on concrete variable names and address displacements without requiring the original structure layout to be fully recovered.

\textbf{Numeric normalization:}
PLC-Bin2Src standardizes common Boolean and null-pointer forms and removes recognized cast syntax before feature extraction. For example, alternative Boolean expressions are converted to common forms, and \texttt{0x02} is normalized to \texttt{2}. During feature extraction, numerical literals are represented by normalized constant labels. The token-level rules retain selected values, including 0, 1, and 2, while grouping other values under a generic constant label; further structural abstraction can merge these labels into a common constant category. This reduces sensitivity to literal formatting and concrete numerical values when comparing operation and dependency patterns.

\subsection{Representation Construction}

{From} {the} normalized function-level representations, PLC-Bin2Src constructs the CDFG and FCG representations to capture intraprocedural and interprocedural structure, respectively. The Symbols representation separately summarizes recovered-symbol and identifier evidence without encoding graph topology.

\subsubsection{CDFG}

The CDFG representation describes control and data semantics within functions. A control-flow graph (CFG) characterizes program execution transitions, whereas a data-flow graph (DFG) describes variable definitions and uses and the propagation of operation results. A CDFG jointly expresses control and data relationships within the same structure~\cite{ferrante1987pdg,cummins2021programl}. PLC-Bin2Src constructs a task-oriented CDFG for binary2source matching. The conditions, operations, calls, variables, and dependencies available after preprocessing are organized around the extracted core functions to form function-level CDFG representations.

PLC-Bin2Src {constructs CDFGs by extracting} semantic events such as conditional decisions, loop control, assignment statements, function calls, memory reads and writes, and arithmetic expressions from the {normalized} C-like representations.

The graph mainly includes three types of nodes:

\begin{itemize}
    \item \emph{Operation nodes:} program events such as conditional branches, loops, assignments, function calls, and returns;
    \item \emph{Arithmetic nodes:} expression operations such as addition, subtraction, multiplication, division, comparison, and logical operations;
    \item \emph{Data nodes:} local variables, memory objects, and constants.
\end{itemize}

Directed edges between nodes record control and data semantics:

\begin{itemize}
    \item Control relationships describe conditional constraints and the local execution order of semantic events within a function;
    \item Data relationships describe propagation from variables or constants to arithmetic nodes, from operation results to subsequent operations, and from return values to target variables.
\end{itemize}

For function calls, the system further recovers call nodes and their argument order from the {syntactic structure of expressions}, connects explicit variable definitions to subsequent uses, and, when determinable, establishes propagation relationships from call return values to temporary variables or return targets. Consecutive calls and their local execution order are also encoded as structural features {to} {retain} call--data relationships recoverable on both the binary and source sides.

After graph construction, PLC-Bin2Src converts the available control, data, and call items into CDFG documents. When statement-level graph structure is recoverable, it performs two iterations of Weisfeiler--Lehman graph kernel (WL) relabeling~\cite{shervashidze2011wl} and extracts first- to third-order token $n$-grams to describe local subgraph patterns and {the order of consecutive operations}. It further records recoverable call-related structures, including call nodes, arguments, definition--use relationships, and call sequences. All CDFG documents are mapped into a common feature space during vectorization to obtain comparable CDFG vector representations.

\begin{figure}[!t]
    \centering
    \includegraphics[width=\columnwidth]{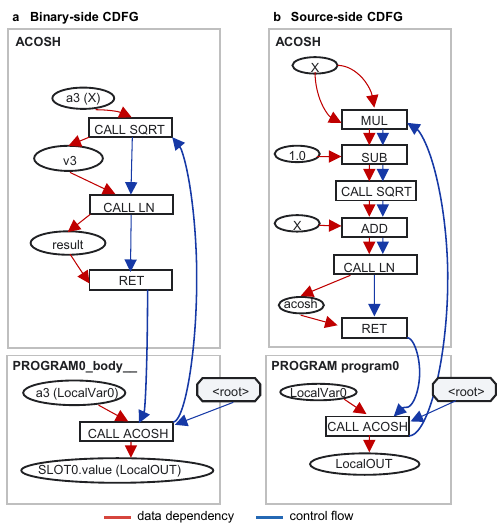}
    \caption{Binary-side and source-side CDFG subgraphs of \texttt{ACOSH}.}
    \label{fig:cdfg_example}
\end{figure}

In Fig.~\ref{fig:cdfg_example}, we select representative nodes and dependency edges to show the local CDFG semantic subgraphs formed around the core computation and function calls on the binary and source sides of \texttt{ACOSH}. Panels (a) and (b) correspond to the binary and source sides, respectively. The source side retains a relatively complete arithmetic hierarchy of the expression, whereas the binary side reflects the call results, temporary variables, and return-value propagation relationships that can be recovered during decompilation. Red edges indicate data dependencies, blue edges indicate control flow or local execution order, and \texttt{<root>} marks the program-entry node.

\subsubsection{FCG}

\begin{figure}[!t]
    \centering
    \includegraphics[width=0.50\columnwidth]{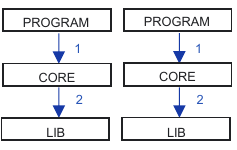}
    \caption{Binary-side and source-side function-call graphs of \texttt{ACOSH}.}
    \label{fig:fcg_example}
\end{figure}

The FCG representation describes {how} {calls} {are organized} among program units. Nodes represent functions, edges represent direct call relationships, and node labels distinguish roles such as program entries, core functions, library functions, and auxiliary functions~\cite{hu2009malware}. Fig.~\ref{fig:fcg_example} presents the call relationships on the binary and source sides of \texttt{ACOSH}. The left and right graphs show the binary and source sides, respectively; \texttt{PROGRAM}{, \texttt{CORE},} and \texttt{LIB} denote the program-entry{, core-function,} and library-function roles{, respectively}, and the numbers denote direct-call counts.

Using the core functions and call information obtained during preprocessing, PLC-Bin2Src constructs an FCG for each normalized binary- and source-side representation. It uses the program entry or extracted core function as the root semantic node and derives structural features from its callees, call relationships, and auxiliary-function roles.

The FCG {encodes} {both} call topology and normalized function roles. Call edges are derived from direct or statically resolved indirect call relationships in the normalized function-level representations. Node labels use unified roles such as INIT, PROGRAM, CORE, LIB, RUNTIME, and HELPER.

\subsubsection{Symbols}

The Symbols representation is an independent lexical summary of function names and related identifiers that can be extracted from source code and binaries. Unlike CDFG and FCG, this representation does not encode operations, dependencies, or call topology. For PLC binaries that have not been stripped or that retain partial symbol clues, {this} evidence can directly support binary2source matching~\cite{xia2023nmatch}.

To construct the Symbols representation, PLC-Bin2Src extracts recovered-symbol labels and related identifiers from the normalized binary and source representations, then maps them to semantic roles using associations provided by preprocessing. It excludes items assigned generic roles such as PROGRAM, INIT, and RUNTIME and generates two feature types, \texttt{NAME\_LABEL} and \texttt{NAME\_ROLE}. Local-variable names and ordinary syntactic identifiers are not used as the primary recovered-symbol evidence.

For example, the binary-side and source-side Symbols documents for \texttt{ACOSH} each contain one occurrence of \texttt{NAME\_LABEL:ACOSH}, which records the recovered function identity, and one occurrence of \texttt{NAME\_ROLE:CORE}, which records its semantic role. In general, the feature counts depend on the recovered evidence available in each document.

\subsubsection{TF-IDF Vectorization}

After extracting features for the three representations, PLC-Bin2Src maps binary and source features into a shared vector space for each representation type. The CDFG feature set consists of structural labels obtained through WL relabeling, $n$-grams of normalized tokens, and sequences of call labels. The FCG feature set records function-role labels and directed caller--callee relationships to describe the local call structure. The Symbols feature set records normalized recovered-symbol names and their associated semantic roles. TF-IDF weighting is applied separately to the features of each representation: it reduces the influence of features {common} across the candidate corpus and {emphasizes} more discriminative features. PLC-Bin2Src therefore obtains the following vectors for binary sample $b$ and source sample $s$:

\begin{equation*}
\mathbf{v}_{\mathrm{cdfg}}^{b},\quad
\mathbf{v}_{\mathrm{fcg}}^{b},\quad
\mathbf{v}_{\mathrm{symbol}}^{b},
\end{equation*}

\begin{equation*}
\mathbf{v}_{\mathrm{cdfg}}^{s},\quad
\mathbf{v}_{\mathrm{fcg}}^{s},\quad
\mathbf{v}_{\mathrm{symbol}}^{s}.
\end{equation*}

{The} {resulting} {binary-side} and {source-side} CDFG, FCG, and Symbols vectors {are comparable}, with each pair defined in the feature space of its representation type.

\subsection{Similarity Calculation}

The similarity-calculation stage computes separate similarity scores for the three representations between a binary query and each source candidate, takes their equal-weight average, and orders the candidates by their fused scores.

\subsubsection{Cosine Similarity}

Given a binary query $b$ and a source candidate $s$, PLC-Bin2Src separately compares their CDFG, FCG, and Symbols vectors. Let the set of representation types be
\begin{equation*}
\mathcal{R}=\{\mathrm{cdfg},\mathrm{fcg},\mathrm{symbol}\}.
\end{equation*}

For any representation type $r\in\mathcal{R}$, the vectors of the binary query and source candidate are denoted by $\mathbf{v}_{r}^{b}$ and $\mathbf{v}_{r}^{s}$, respectively. Their similarity for representation type $r$ is computed using cosine similarity:
\begin{equation}
S_{r}(b,s)=
\frac{\mathbf{v}_{r}^{b}\cdot\mathbf{v}_{r}^{s}}
{\lVert\mathbf{v}_{r}^{b}\rVert_{2}\lVert\mathbf{v}_{r}^{s}\rVert_{2}}.
\label{eq:representation_similarity}
\end{equation}

{The} system {thus} obtains one {independently computed} similarity score for each of the three representations:
\begin{equation*}
S_{\mathrm{cdfg}}(b,s),\quad
S_{\mathrm{fcg}}(b,s),\quad
S_{\mathrm{symbol}}(b,s).
\end{equation*}

The three scores quantify similarity in function-level control--data representations, program-level call representations, and recovered-symbol evidence, respectively, and {serve} {as} inputs to equal-weight fusion.

\subsubsection{Equal-Weight Similarity Fusion}

A single representation captures only one aspect of program semantics. PLC-Bin2Src therefore assigns equal weights to the similarity scores from the three representations, allowing intraprocedural structure, program-level call relationships, and recovered-symbol evidence to jointly determine candidate ranking.

For any binary sample $b$ and source candidate $s$, PLC-Bin2Src computes the final matching score as
\begin{equation}
S_{\mathrm{fused}}(b,s)=\frac{1}{3}\left(
S_{\mathrm{cdfg}}(b,s)+S_{\mathrm{fcg}}(b,s)+S_{\mathrm{symbol}}(b,s)
\right).
\label{eq:fusion}
\end{equation}

Equivalently, the three weights shown in Fig.~\ref{fig:framework} satisfy $w_{\mathrm{cdfg}}=w_{\mathrm{fcg}}=w_{\mathrm{symbol}}=1/3$. 
The resulting scalar score is used for candidate ranking.

\subsubsection{Top-k Candidate Selection}

For a binary sample $b$, PLC-Bin2Src computes its fused score with every ST program in the given source-candidate set $\mathcal{S}_{b}$ and {ranks} {the} {candidates} by sorting the scores in descending order.

Formally, let the source-candidate set of binary query $b$ be $\mathcal{S}_{b}$. PLC-Bin2Src computes the fused score between $b$ and each candidate $s\in\mathcal{S}_{b}$:
\begin{equation*}
S_{\mathrm{fused}}(b,s),\qquad s\in\mathcal{S}_{b}.
\end{equation*}

The candidate sequence is then obtained by sorting the fused scores in descending order:
\begin{equation}
\pi_b=\operatorname{argsort}^{\downarrow}_{s\in\mathcal{S}_{b}}S_{\mathrm{fused}}(b,s).
\label{eq:ranking}
\end{equation}

The position of the true source code in candidate sequence $\pi_b$ is denoted by $\operatorname{rank}(b)$. A smaller value indicates that the correct correspondence {ranks} closer to the top. The first $k$ elements of $\pi_b$ form the top-$k$ candidate set returned by PLC-Bin2Src.

\section{Evaluation}

{We} {evaluate} {the} effectiveness of PLC-Bin2Src, the {contributions} of different representation combinations, and computational efficiency {through} the following research questions:

\begin{itemize}
    \item \textbf{RQ1:} How effective is PLC-Bin2Src in matching PLC binaries compiled by different platforms to their corresponding source files?
    \item \textbf{RQ2:} How do different combinations of the three semantic representations affect binary2source matching performance?
    \item \textbf{RQ3:} What is the runtime cost of each stage of PLC-Bin2Src?
\end{itemize}

\subsection{Experimental Setup}

\subsubsection{Evaluation Dataset}

PLC-BEAD~\cite{achamyeleh2025plcbead} contains 2431 samples. We evaluate 2358 samples for which a unique binary--source correspondence can be established and all three representations can be generated. These samples cover CODESYS, GEB, OpenPLC v2, and OpenPLC v3. All 73 excluded samples are from CODESYS. For 29 samples, the current MATIEC pipeline does not generate valid C-like representations because of unresolved library calls or POU types, compiler-access exceptions, or syntax and naming conflicts. For the remaining 44 samples, we cannot establish a unique, verifiable mapping to a source file. Table~\ref{tab:dataset} reports the resulting platform distribution: 482 CODESYS samples, 617 GEB samples, 619 OpenPLC v2 samples, and 640 OpenPLC v3 samples.

\begin{table}[h]
\caption{Numbers of valid samples used in the experiments.}
\label{tab:dataset}
\centering
\footnotesize
\renewcommand{\arraystretch}{1.08}
\begin{tabular}{@{}c c c@{}}
\toprule
Platform & Binary form & Number of samples \\
\midrule
CODESYS & \texttt{.app} application container & 482 \\
GEB & ARM ELF & 617 \\
OpenPLC v2 & ARM ELF (\texttt{.exe} file) & 619 \\
OpenPLC v3 & ARM ELF (\texttt{.exe} file) & 640 \\
\midrule
Total & -- & 2358 \\
\bottomrule
\end{tabular}
\end{table}

For each toolchain, we construct a compatible candidate corpus containing the corresponding 482, 617, 619, or 640 ST programs. Every source program in a corpus can be compiled into a valid binary by that toolchain and has a known binary correspondence. A source program is not inherently platform specific; if several toolchains compile the same ST program successfully, it can appear in several corpora. For each binary query, PLC-Bin2Src scores every source candidate in the compatible corpus and sorts the candidates by fused score.

\textbf{Implementation:} The binary frontend uses IDA Pro and the Hex-Rays Decompiler to generate C-like representations of target functions~\cite{hexrays}. For OpenPLC wrapper functions that Hex-Rays fails to decompile, a specialized ARM32 lifter uses Capstone~\cite{capstone} disassembly, available symbol information, and predefined wrapper templates to generate approximate C-like representations. The source frontend uses MATIEC to convert ST into a C-style representation~\cite{matiec} and {then} uses a C preprocessor to {expand} {macros}. The system constructs the CDFG, FCG, and Symbols feature documents, computes a similarity score for each representation, and applies fixed equal-weight fusion. The method {requires} {no} training. 


\textbf{Baseline selection:} To the best of our knowledge, no existing approach directly supports binary2source matching between PLC binaries and ST source files. Existing binary2source methods primarily target conventional C/C++ binaries~\cite{miyani2017binpro,yu2020codecmr,gui2022xlir,jiang2024binaryai}, and cannot be directly applied to the PLC artifacts in PLC-BEAD. Therefore, we do not include a direct external baseline. Instead, we use the individual CDFG, FCG, and Symbols representations and their pairwise combinations as controlled internal baselines to evaluate the contribution of each representation and their fusion.

\subsubsection{Evaluation Metrics}

We use the following metrics to evaluate binary2source matching performance:

\begin{itemize}
    \item \textbf{Recall@$k$:} the proportion of queries for which the true source code appears among the top $k$ candidates.
    \item \textbf{Mean reciprocal rank (MRR):} the mean of the reciprocal ranks of the correct {source} {across} all samples, reflecting overall ranking quality.
\end{itemize}

For a query set $\mathcal{B}$, let the position of the true source code in the candidate {list} for query $b$ be $\operatorname{rank}(b)$. Recall@$k$ is defined as
\begin{equation}
\operatorname{Recall@}k=\frac{1}{|\mathcal{B}|}\sum_{b\in\mathcal{B}}
\mathbb{I}\!\left(\operatorname{rank}(b)\leq k\right).
\label{eq:recall_at_k}
\end{equation}

Here, $\mathbb{I}(\cdot)$ is the indicator function. We report results for $k=1$ and $k=5$, which measure the proportions of queries for which the true source code is ranked first and within the top five, respectively.

MRR is defined as
\begin{equation}
\operatorname{MRR}=\frac{1}{|\mathcal{B}|}\sum_{b\in\mathcal{B}}
\frac{1}{\operatorname{rank}(b)}.
\label{eq:mrr}
\end{equation}

Unlike Recall@1 and Recall@5, which consider only whether a match occurs within a specific ranking range, MRR is more sensitive to the exact position of the true source code in the candidate list. When the true source code {ranks} closer to the top, its reciprocal rank and the MRR {increase}.

Recall@1, Recall@5, and MRR evaluate binary2source matching from three perspectives: first-position retrieval, coverage of the leading candidate set, and overall ranking quality, respectively.

\subsection{RQ1: Overall and Platform-Level Matching Results}

We aggregate the ranks of all queries and then {report} results by platform. Table~\ref{tab:overall_results} reports Recall@1, Recall@5, and MRR overall and for each platform.

\begin{table}[h]
\caption{Overall and platform-level matching results.}
\label{tab:overall_results}
\centering
\footnotesize
\setlength{\tabcolsep}{1.0pt}
\renewcommand{\arraystretch}{1.08}
\begin{tabular}{@{}c c c c c c@{}}
\toprule
Platform & \shortstack{Number of\\samples} & Recall@1 & Recall@5 & MRR & \shortstack{Number not\\ranked first} \\
\midrule
CODESYS & 482 & 98.55\% & 99.38\% & 0.9895 & 7 \\
GEB & 617 & 95.79\% & 99.84\% & 0.9770 & 26 \\
OpenPLC v2 & 619 & 94.83\% & 99.68\% & 0.9716 & 32 \\
OpenPLC v3 & 640 & 95.00\% & 99.69\% & 0.9725 & 32 \\
\midrule
Overall & 2358 & 95.89\% & 99.66\% & 0.9769 & 97 \\
\bottomrule
\end{tabular}
\end{table}

On the 2358 valid samples, PLC-Bin2Src achieves 95.89\% Recall@1, 99.66\% Recall@5, and an MRR of 0.9769. Across the four platforms, Recall@1 ranges from 94.83\% to 98.55\%, and Recall@5 ranges from 99.38\% to 99.84\%. CODESYS has the highest Recall@1 and OpenPLC v2 the lowest, with a difference of 3.72 percentage points.

\emph{Answer to RQ1:} PLC-Bin2Src achieves 95.89\% Recall@1 on 2358 queries. Among the 97 queries for which the true source is not ranked first, 89 still place it within the top five.

\subsection{RQ2: Representation Combination Analysis}
\label{sec:rq2}

To compare the evidence provided by different representations, we evaluate each representation individually, all three {pairwise} {combinations}, and the full configuration using all three representations. Similarity scores from the included representations contribute equally: the weight is 1 when using one representation, $1/2$ per score when using two, and $1/3$ per score when using all three. Table~\ref{tab:representation_ablation} reports the results over all 2358 queries.

\begin{table}[h]
\caption{Matching results for different representation combinations.}
\label{tab:representation_ablation}
\centering
\footnotesize
\setlength{\tabcolsep}{4.0pt}
\renewcommand{\arraystretch}{1.08}
\begin{tabular}{@{}c c c c@{}}
\toprule
Representation combination & Recall@1 & Recall@5 & MRR \\
\midrule
CDFG only & 73.11\% & 85.79\% & 0.7902 \\
FCG only & 91.94\% & 98.52\% & 0.9501 \\
Symbols only & 81.89\% & 88.30\% & 0.8483 \\
\midrule
CDFG+FCG & 93.55\% & 98.69\% & 0.9594 \\
CDFG+Symbols & 93.21\% & 99.41\% & 0.9601 \\
FCG+Symbols & 93.77\% & 99.24\% & 0.9633 \\
\midrule
PLC-Bin2Src & 95.89\% & 99.66\% & 0.9769 \\
\bottomrule
\end{tabular}
\end{table}

Among the individual representations, FCG performs best, with 91.94\% Recall@1. Equal pairwise fusion yields 93.55\% Recall@1 for CDFG+FCG, 93.21\% for CDFG+Symbols, and 93.77\% for FCG+Symbols. The full configuration with equal weights for all three similarity scores reaches 95.89\% Recall@1, 99.66\% Recall@5, and an MRR of 0.9769, improving on the strongest pairwise combination by 2.12 percentage points, 0.42 percentage points, and 0.0136, respectively. These results show that function-level control--data representations, program-level call representations, and topology-independent symbol evidence provide complementary signals for candidate ranking.

\emph{Answer to RQ2:} Fixed equal-weight fusion of the similarities from all three representations outperforms every individual representation and every equal-weight pairwise combination, achieving 95.89\% Recall@1, 99.66\% Recall@5, and an MRR of 0.9769.

\subsection{RQ3: Runtime Cost by Stage}
\label{sec:rq3_runtime}

We report runtime for three pipeline stages. \emph{Preprocessing} covers the source and binary frontends, \emph{representation construction} builds the CDFG, FCG, and Symbols representations {and performs TF-IDF vectorization}, and \emph{similarity calculation} {includes cosine similarity}, equal-weight fusion, and candidate ranking.

\begin{figure}[h]
    \centering
    \includegraphics[width=\columnwidth]{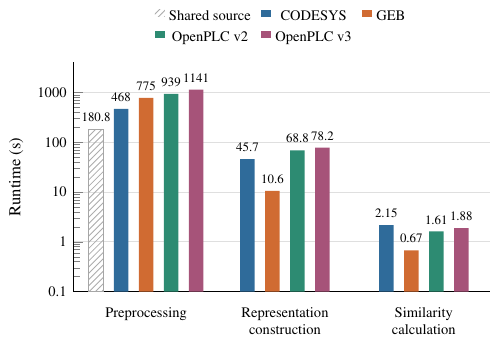}
    \caption{Runtime evaluation by pipeline stage.}
    \label{fig:runtime_stage_cost}
\end{figure}

Figure~\ref{fig:runtime_stage_cost} reports total runtime in seconds on a logarithmic scale. The source repository is preprocessed once in 180.81~s. Across the four platforms, binary preprocessing requires 467.73--1141.17~s and dominates the total cost, whereas representation construction requires {10.56--78.24}~s and similarity calculation requires only {0.67--2.15}~s.

\emph{Answer to RQ3:} Once preprocessing and representation construction are complete, PLC-Bin2Src performs full-corpus binary2source matching in only {0.67--2.15}~s across the four platforms.

\subsection{Analysis of Results Not Ranked First}

Among the 97 queries for which the true source is not ranked first, 89 still place it within the top five: 73 rank it second and 16 rank it third through fifth. The remaining eight place it outside the top five. The median score difference between the first-ranked incorrect candidate and the true source is 0.00975. Most errors therefore occur among leading candidates with similar scores, although a small number place the true source substantially lower.

Four representative cases illustrate these errors. For the GEB query \texttt{FT\_TN16}, \texttt{FT\_TN8} is ranked first and the true source is ranked third. For \texttt{STACK\_32} on OpenPLC v2, \texttt{STACK\_16} is ranked first and the true source is ranked second; their FCG and Symbols scores are identical, while their CDFG scores differ by only 0.000028. In both cases, name normalization removes the numeric suffixes that distinguish the variants, making their normalized names identical and weakening the naming evidence available for matching. For the CODESYS query \texttt{\_ARRAY\_ADD}, the recovered symbol slice is incorrectly associated with \texttt{\_ARRAY\_ABS}, causing all name-bearing representations to favor that candidate and placing the true source at rank 216. Finally, for the GEB query \texttt{URL\_TO\_STRING}, \texttt{\_RMP\_B} is ranked first and the true source is ranked 305th. In this case, no symbol is recovered, and neither the CDFG nor the FCG representation provides sufficiently discriminative features.

{These} cases {show} three recurring patterns. First, variants within the same function family can receive identical or nearly equal similarity scores for the individual representations. Second, differences in retained types, constants, or intermediate operations may leave too little fine-grained evidence to change the ranking. Third, inaccurate frontend associations can distort the resulting evidence, as in \texttt{\_ARRAY\_ADD}. Similar errors occur in OpenPLC v2 and v3, suggesting that related compilation toolchains preserve similar features for some function variants.

\section{Discussion}

The evaluated binaries retain recoverable naming clues {from} the PLC compilation and packaging toolchains represented in PLC-BEAD{.} PLC-Bin2Src does not introduce these names from the source code. In GEB and OpenPLC, the binaries expose compiler-generated POU and function names, whereas CODESYS \texttt{.app} containers preserve binary-visible names and pointers in metadata records. PLC-BinX reports the same platform behavior and associates the recovered names with function-level representations{.} {We} therefore {follow} PLC-BEAD and PLC-BinX in analyzing these engineering-generated compiled artifacts~\cite{achamyeleh2025plcbead,jia2026plcbinx}. This use of information recoverable from compiled PLC artifacts is also consistent with prior PLC reverse-engineering studies~\cite{keliris2019icsref,benkraouda2023plcspecific}.

Recoverable symbols are useful but incomplete matching evidence. As shown in Table~\ref{tab:representation_ablation}, the Symbols-only configuration achieves 81.89\% Recall@1, whereas CDFG+Symbols and FCG+Symbols reach 93.21\% and 93.77\%, respectively{.} {Equal-weight} fusion of the similarities from all three representations reaches 95.89\%. Thus, even when naming clues are widely available, they do not uniquely determine source correspondence. CDFG and FCG contribute complementary control--data and call-organization evidence that {distinguishes among} candidates sharing similar names or function roles. PLC-Bin2Src therefore uses recovered symbols as {a} complementary representation rather than reducing binary2source matching to name comparison, consistent with naming-aware binary-similarity research~\cite{xia2023nmatch}.

\section{Threats to Validity}

\subsection{Construct Validity}

The {experiments inherit their} ground-truth correspondences from the data organization of PLC-BEAD~\cite{achamyeleh2025plcbead}. {Any errors in} the source--binary correspondences, platform labels, or compilation artifacts {of} the original samples will also enter the evaluation. 

\subsection{Internal Validity}

Core-function extraction, decompilation, ST-to-C conversion, semantic normalization, and feature extraction depend on multiple analysis tools and rules. IDA Pro/Hex-Rays decompilation results, Capstone disassembly, MATIEC code generation, and the {project's} parsing and normalization processes may all produce omissions or representation bias~\cite{matiec,hexrays,capstone,andriesse2016disassembly,pei2021xda,gao2025decompilebench}. The experiments use consistent frontend processes, candidate sets, and evaluation rules for all query corpora to reduce differences in processing conditions, but tool errors cannot be eliminated. 

\subsection{External Validity}

This paper evaluates CODESYS, GEB, OpenPLC v2, and OpenPLC v3, but these platforms do not cover all PLC vendors, compilation toolchains, architectures, and optimization configurations. The evaluation follows PLC-BEAD and PLC-BinX in using engineering-generated compiled artifacts from the covered toolchains, which retain recoverable naming clues~\cite{achamyeleh2025plcbead,jia2026plcbinx}. The conclusions therefore concern these toolchains and artifact types; binaries subjected to deliberate symbol stripping or name obfuscation constitute additional conditions for future evaluation. The fixed equal-weight fusion rule {may} also not be optimal {when} the quality of the extracted representations {in other datasets or platform distributions} differs substantially from that in PLC-BEAD.

\section{Related Work}

\subsection{General-Purpose Binary2Source Matching}

Binary2source matching identifies correspondences between binary artifacts and source code, supporting software composition analysis, code-reuse detection, and provenance tracing. Existing approaches can be broadly grouped by their principal matching evidence into string-based, structure-based, and learning-based methods.

\emph{String-based methods.} These methods exploit strings, identifiers, constants, and other lexical features that can be extracted from both source code and binaries. BAT combines strings with compression similarity and binary differencing to identify code reuse in firmware~\cite{hemel2011license}. OSSPolice uses string and function features to detect open-source components in mobile applications~\cite{duan2017osspolice}, while B2SFinder matches source and binary feature instances for open-source reuse detection~\cite{feng2019b2sfinder}. B2SMatcher uses string literals and exported function names for coarse matching and supplements them with function-level constants and call-graph degrees for fine-grained version identification~\cite{ban2021b2smatcher}.

\emph{Structure-based methods.} These methods compare relations among program elements to reduce dependence on exact lexical correspondence. BinPro combines constants, library calls, function-call relationships, and iterative bipartite function matching to determine binary provenance~\cite{miyani2017binpro}. BugGraph, GraphBinMatch, and BSMDG use attributed control-flow graphs, call relationships, or decompiler-derived graph structures to preserve program semantics across compilation~\cite{ji2021buggraph,tehranijamsaz2024graphbinmatch,aljebreen2025bsmdg}. Their designs are related to broader graph representations, including program dependence graphs~\cite{ferrante1987pdg}, Weisfeiler--Lehman neighborhood aggregation~\cite{shervashidze2011wl}, and ProGraML's joint representation of control, data, and call relations~\cite{cummins2021programl}. Function-call graphs and recoverable names can provide additional program-level evidence~\cite{hu2009malware,xia2023nmatch}.

\emph{Learning-based methods.} These methods learn a shared representation in which corresponding source and binary code are close. CodeCMR uses a text model and a graph neural network to encode source and binary functions, respectively~\cite{yu2020codecmr}. XLIR reduces cross-language and cross-representation differences through intermediate representations~\cite{gui2022xlir}, while CrossCode2Vec combines abstract-syntax paths with hierarchical sequence models to learn cross-representation features~\cite{yu2025crosscode2vec}. BinaryAI combines transformer-based function embeddings with link-time locality for binary software composition analysis~\cite{jiang2024binaryai}. Recent work also uses strings, constants, and external calls as anchors before reranking source functions with a large language model~\cite{gagnon2026recovery}. Related learning-based binary-similarity methods, including Gemini, DeepBinDiff, Order Matters, and Trex, further demonstrate the value of structural and execution-semantic representations across compilation~\cite{xu2017gemini,duan2020deepbindiff,yu2020order,pei2023trex}.

\subsection{PLC Binary Analysis}

PLC binary analysis must account for platform-specific compilation toolchains, executable formats, and runtime organizations. ICSREF identifies domain-specific features in CODESYS binaries to recover function boundaries, call relationships, and control flow~\cite{keliris2019icsref}. Benkraouda et al. show that programming language, compiler version, target architecture, and PLC vendor all affect binary representations~\cite{benkraouda2023plcspecific}. Other studies reconstruct PLC control logic, recover program structures, or decompile proprietary artifacts by exploiting platform and runtime characteristics~\cite{qasim2019reconstruction,qasim2020forensics,sang2024decompile,geng2024control}.

Recovered PLC structures also support downstream security analysis. ICSFuzz and ICS-QUARTZ use binary-level control information for instrumented and scan-cycle-aware fuzzing, while ICSPatch derives data-dependence information for vulnerability localization and runtime hotpatching~\cite{tychalas2021icsfuzz,villa2025icsquartz,rajput2023icspatch}. PLC-BEAD provides 2431 paired binaries and ST programs spanning CODESYS, GEB, OpenPLC v2, and OpenPLC v3~\cite{achamyeleh2025plcbead}. Built on this corpus, PLC-BinX supports cross-platform core-function identification and function-level semantic recovery~\cite{jia2026plcbinx}.

The two bodies of work address complementary aspects of PLC binary2source matching. Existing binary2source methods are mainly designed for general-purpose software and conventional executable or decompiler representations. Existing PLC binary-analysis methods recover and analyze semantics from platform-specific artifacts but do not identify the corresponding ST source file from a candidate repository. PLC-Bin2Src connects these directions through platform-aware semantic recovery, aligned binary and source representations, and source-candidate ranking.

\section{Conclusion}

This paper presents PLC-Bin2Src for retrieving {the} ST source files {corresponding to} binary artifacts from CODESYS, GEB, OpenPLC v2, and OpenPLC v3. PLC-Bin2Src constructs comparable CDFG, FCG, and recovered-symbol representations and combines their {similarity scores} for candidate ranking. On PLC-BEAD, it achieves 95.89\% Recall@1, 99.66\% Recall@5, and an MRR of 0.9769 across the four platforms. These results demonstrate the feasibility of cross-platform binary2source matching for opaque PLC binaries and provide a foundation for broader provenance and composition analysis.

\bibliographystyle{IEEEtran}
\bibliography{references}

@inproceedings{lopez2024sok,
	author = {Efr{\'e}n L{\'o}pez-Morales and Ulysse Planta and Carlos Rubio-Medrano and Ali Abbasi and Alvaro A. Cardenas},
  title     = {{SoK}: Security of Programmable Logic Controllers},
  booktitle = {33rd USENIX Security Symposium (USENIX Security 24)},
  year      = {2024},
  pages     = {7103--7122},
  url       = {https://www.usenix.org/conference/usenixsecurity24/presentation/lopez-morales}
}

@inproceedings{keliris2019icsref,
  author    = {Anastasis Keliris and Michail Maniatakos},
  title     = {{ICSREF}: A Framework for Automated Reverse Engineering of Industrial Control Systems Binaries},
  booktitle = {Network and Distributed System Security Symposium (NDSS)},
  year      = {2019},
  doi       = {10.14722/ndss.2019.23271}
}

@inproceedings{tychalas2021icsfuzz,
  author = {Dimitrios Tychalas and Hadjer Benkraouda and Michail Maniatakos},
  title     = {{ICSFuzz}: Manipulating {I/Os} and Repurposing Binary Code to Enable Instrumented Fuzzing in {ICS} Control Applications},
  booktitle = {30th USENIX Security Symposium (USENIX Security 21)},
  pages     = {2847--2862},
  year      = {2021},
  url       = {https://www.usenix.org/conference/usenixsecurity21/presentation/tychalas}
}

@inproceedings{rajput2023icspatch,
  author = {Prashant Hari Narayan Rajput and Constantine Doumanidis and Michail Maniatakos},
  title     = {{ICSPatch}: Automated Vulnerability Localization and Non-Intrusive Hotpatching in Industrial Control Systems Using Data Dependence Graphs},
  booktitle = {32nd USENIX Security Symposium (USENIX Security 23)},
  pages     = {6861--6876},
  year      = {2023},
  url       = {https://www.usenix.org/conference/usenixsecurity23/presentation/rajput}
}

@misc{jia2026plcbinx,
  author       = {A. Jia and Y. Duan and H. Jiang and Z. Tian and Z. Ren and X. Li},
  title        = {{PLC-BinX}: A Cross-Platform Binary Code Analysis Framework for {PLC} Binaries},
  year         = {2026},
  howpublished = {arXiv:2605.17392},
  url          = {https://arxiv.org/abs/2605.17392}
}

@inproceedings{hemel2011license,
  author    = {Armijn Hemel and Karl Trygve Kalleberg and Rob Vermaas and Eelco Dolstra},
  title     = {Finding Software License Violations through Binary Code Clone Detection},
  booktitle = {8th Working Conference on Mining Software Repositories},
  pages     = {63--72},
  year      = {2011},
  doi       = {10.1145/1985441.1985453}
}

@inproceedings{duan2017osspolice,
  author    = {Ruian Duan and Ashish Bijlani and Meng Xu and Taesoo Kim and Wenke Lee},
  title     = {Identifying Open-Source License Violation and 1-Day Security Risk at Large Scale},
  booktitle = {2017 ACM SIGSAC Conference on Computer and Communications Security},
  pages     = {2169--2185},
  year      = {2017},
  doi       = {10.1145/3133956.3134048}
}

@misc{miyani2017binpro,
  author       = {Dhaval Miyani and Zhen Huang and David Lie},
  title        = {{BinPro}: A Tool for Binary Source Code Provenance},
  year         = {2017},
  howpublished = {arXiv:1711.00830},
  url          = {https://arxiv.org/abs/1711.00830}
}

@inproceedings{feng2019b2sfinder,
  author    = {Feng, Muyue and Yuan, Zimu and Li, Feng and Ban, Gu and Xiao, Yang and Wang, Shiyang and Tang, Qian and Su, He and Yu, Chendong and Xu, Jiahuan and Piao, Aihua and Xue, Jingling and Huo, Wei},
  title     = {{B2SFinder}: Detecting Open-Source Software Reuse in {COTS} Software},
  booktitle = {34th IEEE/ACM International Conference on Automated Software Engineering (ASE)},
  pages     = {1038--1049},
  year      = {2019},
  doi       = {10.1109/ASE.2019.00100}
}

@article{ban2021b2smatcher,
  author = {Gu Ban and Lili Xu and Yang Xiao and Xinhua Li and Zimu Yuan and Wei Huo},
  title   = {{B2SMatcher}: Fine-Grained Version Identification of Open-Source Software in Binary Files},
  journal = {Cybersecurity},
  volume  = {4},
  year    = {2021},
  doi     = {10.1186/s42400-021-00085-7},
  note    = {Art. no. 21}
}

@inproceedings{yu2020codecmr,
  author    = {Z. Yu and W. Zheng and J. Wang and Q. Tang and S. Nie and S. Wu},
  title     = {{CodeCMR}: Cross-Modal Retrieval for Function-Level Binary Source Code Matching},
  booktitle = {Advances in Neural Information Processing Systems 33},
  pages     = {3872--3883},
  year      = {2020},
  url       = {https://papers.nips.cc/paper/2020/hash/285f89b802bcb2651801455c86d78f2a-Abstract.html}
}

@inproceedings{gui2022xlir,
  author    = {Gui, Yi and Wan, Yao and Zhang, Hongyu and Huang, Huifang and Sui, Yulei and Xu, Guandong and Shao, Zhiyuan and Jin, Hai},
  title     = {Cross-Language Binary-Source Code Matching with Intermediate Representations},
  booktitle = {IEEE International Conference on Software Analysis, Evolution and Reengineering (SANER)},
  pages     = {601--612},
  year      = {2022},
  doi       = {10.1109/SANER53432.2022.00077}
}

@article{yu2025crosscode2vec,
  author  = {G. Yu and J. An and J. Lyu and W. Huang and W. Fan and Y. Cheng and A. Sui},
  title   = {{CrossCode2Vec}: A Unified Representation across Source and Binary Functions for Code Similarity Detection},
  journal = {Neurocomputing},
  volume  = {620},
  year    = {2025},
  doi     = {10.1016/j.neucom.2024.129238},
  note    = {Art. no. 129238}
}

@inproceedings{benkraouda2023plcspecific,
  author = {Benkraouda, Hadjer and Agrawal, Anand and Tychalas, Dimitrios and Sazos, Marios and Maniatakos, Michail},
  title     = {Towards {PLC}-Specific Binary Analysis Tools: An Investigation of {CODESYS}-Compiled {PLC} Software Applications},
  booktitle = {5th Workshop on CPS and IoT Security and Privacy},
  pages     = {83--89},
  year      = {2023},
  doi       = {10.1145/3605758.3623499}
}

@inproceedings{achamyeleh2025plcbead,
  author    = {Y. G. Achamyeleh and S. Y. Yu and G. Q. Araya and M. A. Al Faruque},
  title     = {Bridging the Binary Analysis Gap: A Cross-Compiler Dataset and Neural Framework for Industrial Control Systems},
  booktitle = {31st ACM SIGKDD Conference on Knowledge Discovery and Data Mining V.2},
  pages     = {5260--5269},
  year      = {2025},
  doi       = {10.1145/3711896.3737373}
}

@standard{iec61131,
  author       = {{International Electrotechnical Commission}},
  title        = {{IEC 61131-3:2013, Programmable Controllers---Part 3: Programming Languages}},
  organization = {IEC},
  address      = {Geneva, Switzerland},
  edition      = {3},
  year         = {2013}
}

@misc{matiec,
  author       = {{MATIEC Project}},
  title        = {{MATIEC}: {IEC} 61131-3 Compiler},
  howpublished = {Software documentation},
  url          = {https://openplcproject.gitlab.io/matiec/},
  note         = {Accessed: Aug. 26, 2026}
}

@article{ferrante1987pdg,
  author  = {Jeanne Ferrante and Karl J. Ottenstein and Joe D. Warren},
  title   = {The Program Dependence Graph and Its Use in Optimization},
  journal = {ACM Transactions on Programming Languages and Systems},
  volume  = {9},
  number  = {3},
  pages   = {319--349},
  year    = {1987},
  doi     = {10.1145/24039.24041}
}

@inproceedings{cummins2021programl,
  author    = {Chris Cummins and Zacharias V. Fisches and Tal Ben-Nun and Torsten Hoefler and Michael F. P. O'Boyle and Hugh Leather},
  title     = {{ProGraML}: A Graph-Based Program Representation for Data Flow Analysis and Compiler Optimizations},
  booktitle = {38th International Conference on Machine Learning},
  series    = {Proceedings of Machine Learning Research},
  volume    = {139},
  pages     = {2244--2253},
  year      = {2021},
  url       = {https://proceedings.mlr.press/v139/cummins21a.html}
}

@inproceedings{hu2009malware,
  author    = {Hu, Xin and Chiueh, Tzi-cker and Shin, Kang G.},
  title     = {Large-Scale Malware Indexing Using Function-Call Graphs},
  booktitle = {16th ACM Conference on Computer and Communications Security},
  pages     = {611--620},
  year      = {2009},
  doi       = {10.1145/1653662.1653736}
}

@article{xia2023nmatch,
  author  = {B. Xia and J. Pang and X. Zhou and Z. Shan and J. Wang and F. Yue},
  title   = {Binary Code Similarity Analysis Based on Naming Function and Common Vector Space},
  journal = {Scientific Reports},
  volume  = {13},
  number  = {1},
  year    = {2023},
  doi     = {10.1038/s41598-023-42769-9},
  note    = {Art. no. 15676}
}

@article{shervashidze2011wl,
  author  = {Nino Shervashidze and Pascal Schweitzer and Erik Jan van Leeuwen and Kurt Mehlhorn and Karsten M. Borgwardt},
  title   = {Weisfeiler-Lehman Graph Kernels},
  journal = {Journal of Machine Learning Research},
  volume  = {12},
  pages   = {2539--2561},
  year    = {2011},
  url     = {https://www.jmlr.org/papers/v12/shervashidze11a.html}
}

@misc{hexrays,
  author       = {{Hex-Rays}},
  title        = {{IDA Pro} Disassembler and Debugger},
  howpublished = {Computer software},
  url          = {https://hex-rays.com/ida-pro/},
  note         = {Accessed: Aug. 26, 2026}
}

@misc{capstone,
  author       = {{Capstone Project}},
  title        = {Capstone Disassembly Framework},
  howpublished = {Computer software},
  url          = {https://www.capstone-engine.org/},
  note         = {Accessed: Aug. 26, 2026}
}

@inproceedings{andriesse2016disassembly,
  author    = {Dennis Andriesse and Xi Chen and Victor van der Veen and Asia Slowinska and Herbert Bos},
  title     = {An In-Depth Analysis of Disassembly on Full-Scale x86/x64 Binaries},
  booktitle = {25th USENIX Security Symposium},
  pages     = {583--600},
  year      = {2016},
  url       = {https://www.usenix.org/conference/usenixsecurity16/technical-sessions/presentation/andriesse}
}

@inproceedings{pei2021xda,
  author    = {Kexin Pei and Jonas Guan and David Williams-King and Junfeng Yang and Suman Jana},
  title     = {{XDA}: Accurate, Robust Disassembly with Transfer Learning},
  booktitle = {Network and Distributed System Security Symposium (NDSS)},
  year      = {2021},
  doi       = {10.14722/ndss.2021.23112}
}

@inproceedings{gao2025decompilebench,
  author    = {Z. Gao and Y. Cui and H. Wang and S. Qin and Y. Wang and Z. Bolun and C. Zhang},
  title     = {{DecompileBench}: A Comprehensive Benchmark for Evaluating Decompilers in Real-World Scenarios},
  booktitle = {Findings of the Association for Computational Linguistics: ACL 2025},
  pages     = {23250--23267},
  year      = {2025},
  doi       = {10.18653/v1/2025.findings-acl.1194}
}

@incollection{qasim2019reconstruction,
  author    = {Qasim, Syed Ali and Lopez, Jr., Juan and Ahmed, Irfan},
  title     = {Automated Reconstruction of Control Logic for Programmable Logic Controller Forensics},
  booktitle = {Information Security, ISC 2019},
  publisher = {Springer},
  pages     = {402--422},
  year      = {2019},
  doi       = {10.1007/978-3-030-30215-3_20}
}

@article{qasim2020forensics,
  author  = {Syed Ali Qasim and Jared M. Smith and Irfan Ahmed},
  title   = {Control Logic Forensics Framework Using Built-in Decompiler of Engineering Software in Industrial Control Systems},
  journal = {Forensic Science International: Digital Investigation},
  volume  = {33},
  year    = {2020},
  doi     = {10.1016/j.fsidi.2020.301013},
  note    = {Art. no. 301013}
}

@article{sang2024decompile,
  author  = {C. Sang and J. Wu and J. Li and M. Guizani},
  title   = {From Control Application to Control Logic: {PLC} Decompile Framework for Industrial Control System},
  journal = {IEEE Transactions on Information Forensics and Security},
  volume  = {19},
  pages   = {8685--8700},
  year    = {2024},
  doi     = {10.1109/TIFS.2024.3402117}
}

@article{geng2024control,
  author  = {Y. Geng and X. Che and R. Ma and Q. Wei and M. Wang and Y. Chen},
  title   = {Control Logic Attack Detection and Forensics Through Reverse-Engineering and Verifying {PLC} Control Applications},
  journal = {IEEE Internet of Things Journal},
  volume  = {11},
  number  = {5},
  pages   = {8386--8400},
  year    = {2024},
  doi     = {10.1109/JIOT.2023.3318988}
}

@inproceedings{villa2025icsquartz,
  author    = {C. Villa and C. Doumanidis and H. Lamri and P. H. N. Rajput and M. Maniatakos},
  title     = {{ICS-QUARTZ}: Scan Cycle-Aware and Vendor-Agnostic Fuzzing for Industrial Control Systems},
  booktitle = {Network and Distributed System Security Symposium (NDSS)},
  year      = {2025},
  doi       = {10.14722/ndss.2025.240795}
}

@inproceedings{jiang2024binaryai,
  author    = {L. Jiang and J. An and H. Huang and Q. Tang and S. Nie and S. Wu and Y. Zhang},
  title     = {{BinaryAI}: Binary Software Composition Analysis via Intelligent Binary Source Code Matching},
  booktitle = {IEEE/ACM 46th International Conference on Software Engineering},
  pages     = {1--13},
  year      = {2024},
  doi       = {10.1145/3597503.3639100}
}

@inproceedings{ji2021buggraph,
  author    = {Y. Ji and L. Cui and H. H. Huang},
  title     = {{BugGraph}: Differentiating Source-Binary Code Similarity with Graph Triplet-Loss Network},
  booktitle = {2021 ACM Asia Conference on Computer and Communications Security},
  pages     = {702--715},
  year      = {2021},
  doi       = {10.1145/3433210.3437533}
}

@inproceedings{tehranijamsaz2024graphbinmatch,
  author    = {TehraniJamsaz, Ali and Chen, Hanze and Jannesari, Ali},
  title     = {{GraphBinMatch}: Graph-Based Similarity Learning for Cross-Language Binary and Source Code Matching},
  booktitle = {2024 IEEE International Parallel and Distributed Processing Symposium Workshops (IPDPSW)},
  pages     = {506--515},
  year      = {2024},
  doi       = {10.1109/IPDPSW63119.2024.00103}
}

@article{aljebreen2025bsmdg,
  author  = {G. Aljebreen and R. Alnanih and F. Eassa and M. Khemakhem and K. Jambi and M. U. Ashraf},
  title   = {Binary--Source Code Matching Based on Decompilation Techniques and Graph Analysis},
  journal = {International Journal of Advanced Computer Science and Applications},
  volume  = {16},
  number  = {5},
  year    = {2025},
  doi     = {10.14569/IJACSA.2025.0160525}
}

@misc{gagnon2026recovery,
  author       = {C. E. Gagnon and S. H. H. Ding and P. Charland and B. C. M. Fung},
  title        = {Practical Source Code Recovery from Binary Functions Using Anchor-Based Retrieval and {LLM} Reasoning},
  year         = {2026},
  howpublished = {arXiv:2607.09452},
  url          = {https://arxiv.org/abs/2607.09452}
}

@inproceedings{xu2017gemini,
  author    = {Xiaojun Xu and Chang Liu and Qian Feng and Heng Yin and Le Song and Dawn Song},
  title     = {Neural Network-Based Graph Embedding for Cross-Platform Binary Code Similarity Detection},
  booktitle = {2017 ACM SIGSAC Conference on Computer and Communications Security},
  pages     = {363--376},
  year      = {2017},
  doi       = {10.1145/3133956.3134018}
}

@inproceedings{duan2020deepbindiff,
  author    = {Yue Duan and Xuezixiang Li and Jinghan Wang and Heng Yin},
  title     = {{DeepBinDiff}: Learning Program-Wide Code Representations for Binary Diffing},
  booktitle = {Network and Distributed System Security Symposium (NDSS)},
  year      = {2020},
  doi       = {10.14722/ndss.2020.24311}
}

@article{yu2020order,
  author  = {Z. Yu and R. Cao and Q. Tang and S. Nie and J. Huang and S. Wu},
  title   = {Order Matters: Semantic-Aware Neural Networks for Binary Code Similarity Detection},
  journal = {Proceedings of the AAAI Conference on Artificial Intelligence},
  volume  = {34},
  number  = {1},
  pages   = {1145--1152},
  year    = {2020},
  doi     = {10.1609/aaai.v34i01.5466}
}

@article{pei2023trex,
  author  = {Kexin Pei and Zhou Xuan and Junfeng Yang and Suman Jana and Baishakhi Ray},
  title   = {Learning Approximate Execution Semantics from Traces for Binary Function Similarity},
  journal = {IEEE Transactions on Software Engineering},
  volume  = {49},
  number  = {4},
  pages   = {2776--2790},
  year    = {2023},
  doi     = {10.1109/TSE.2022.3231621}
}

@misc{codesysCompiledLibraries,
  author       = {{CODESYS GmbH}},
  title        = {Command: Save Project as Compiled Library},
  howpublished = {CODESYS Online Help. [Online]. Available:
                  \url{https://content.helpme-codesys.com/en/CODESYS%20Development%20System/_cds_cmd_save_project_as_compiled_library.html}},
  note         = {Accessed: Sep. 5, 2026}
}

@misc{codesysLibraryDeployment,
  author       = {{CODESYS GmbH}},
  title        = {Deployment and Licensing},
  howpublished = {CODESYS Library Development Summary. [Online]. Available: \url{https://content.helpme-codesys.com/en/LibDevSummary/deployment.html}},
  note         = {Accessed: Aug. 30, 2026}
}

\end{document}